\documentclass[sigconf,screen,authordraft,nonacm,review=false,timestamp=false]{acmart}

\AtBeginDocument{%
  }

\copyrightyear{2026}
\acmYear{2026}
\setcopyright{cc}
\setcctype{by}
\acmConference[CHI '26]{Proceedings of the 2026 CHI Conference on Human Factors in Computing Systems}{April 13--17, 2026}{Barcelona, Spain}
\acmBooktitle{Proceedings of the 2026 CHI Conference on Human Factors in Computing Systems (CHI '26), April 13--17, 2026, Barcelona, Spain}
\acmDOI{10.1145/3772318.3791252}
\acmISBN{979-8-4007-2278-3/2026/04}

\usepackage{subcaption}
\usepackage{colortbl}
\usepackage{xcolor}
\usepackage{enumitem}
\usepackage{xspace}

\newcommand{\eg}{e.g.\@\xspace}

\begin{document}

\title[Unraveling Entangled Feeds]{Unraveling Entangled Feeds: Rethinking Social Media Design to Enhance User Well-being}

\author{Ashlee Milton}
\email{milto064@umn.edu}
\orcid{0000-0002-0320-6122}
\affiliation{%
  \institution{University of Minnesota - GroupLens}
  \city{Minneapolis}
  \state{Minnesota}
  \country{USA}
}

\author{Dan Runningen}
\email{runni028@umn.edu}
\orcid{}
\affiliation{%
  \institution{University of Minnesota - GroupLens}
  \city{Minneapolis}
  \state{Minnesota}
  \country{USA}
}

\author{Loren Terveen}
\email{terveen@umn.edu}
\orcid{}
\affiliation{%
  \institution{University of Minnesota - GroupLens}
  \city{Minneapolis}
  \state{Minnesota}
  \country{USA}
}

\author{Harmanpreet Kaur}
\email{harmank@umn.edu}
\orcid{}
\affiliation{%
  \institution{University of Minnesota - GroupLens}
  \city{Minneapolis}
  \state{Minnesota}
  \country{USA}
}

\author{Stevie Chancellor}
\email{steviec@umn.edu}
\orcid{0000-0003-0620-0903}
\affiliation{%
  \institution{University of Minnesota - GroupLens}
  \city{Minneapolis}
  \state{Minnesota}
  \country{USA}
}

\renewcommand{\shortauthors}{Milton et al.}

\begin{abstract}
Social media platforms have rapidly adopted algorithmic curation with little consideration for the potential harm to users' mental well-being. We present findings from design workshops with 21 participants diagnosed with mental illness about their interactions with social media platforms. We find that users develop cause-and-effect explanations, or folk theories, to understand their experiences with algorithmic curation. These folk theories highlight a breakdown in algorithmic design that we explain using the framework of \textit{entanglement}, a phenomenon where there is a disconnect between users' actions and platform outcomes on an emotional level. Participants' designs to address entanglement and mitigate harms centered on contextualizing their engagement and restoring explicit user control on social media. The conceptualization of entanglement and the resulting design recommendations have implications for social computing and recommender systems research, particularly in evaluating and designing social media platforms that support users' mental well-being.
\end{abstract}



\maketitle

\section{Introduction}
Algorithmic curation has transformed social media platforms, like Instagram and TikTok, to deliver personalized content to billions of users~\cite{forbes_stats}. Powered by recommender systems, algorithmic curation provides users with content from anywhere on the platform~\cite{shin2022algorithm} and serves as the primary distribution mechanism for much of the content on current social media platforms. Prior work demonstrates that algorithmic curation has positive effects on users, including information discovery~\cite{milton2024seeking}, identity and self-conceptualization~\cite{lee2022algorithmic,karizat2021algorithmic}, and social connection~\cite{milton2023see,cullen2025not}. Furthermore, platforms can directly improve mental well-being for users with mental illness by providing coping strategies and validation~\cite{yan2014feeling,basch2022deconstructing}, as well as information, community, and support~\cite{pretorius2019young,de2014mental,milton2024seeking}. 

However, concerns are growing that social media platforms and algorithms can harm people and their mental health~\cite{haidt2024anxious,naslund2019risks,karim2020social}. TikTok and Instagram have received widespread public attention for their potential impact on teens and people with mental illness~\cite{guardian_paul,cnn_yurkevich,Healthline}. Prior academic work has also found that social media platforms propagate harmful information about mental health, including dangerous advice about eating disorders~\cite{chancellor2016thyghgapp,chancellor2017multimodal}, propagating challenges that promote self-harm~\cite{milton2022users}, and even triggering or exacerbating dangerous behaviors~\cite{dyson2016systematic,feuston2019everyday}. Disengaging from social media is not viable for many users with mental illness, as social media is their best source of mental health support in the face of societal stigma and limited resources~\cite{naslund2019risks,aguirre2020barriers}.

A growing body of work suggests that social media design features, particularly algorithmic content feeds, may drive harmful experiences for users with mental illnesses. Moreover, several recent HCI studies show that users identified the lack of control over algorithmically curated content as a primary source of harm to their well-being~\cite{milton2023see,milton2024seeking,simpson2022tame}. Participants used metaphors like ``dopamine slot machine,''~\cite{milton2024seeking} ``runaway train,''~\cite{milton2023see} and ``domesticating'' the feed~\cite{simpson2022tame} to describe their fraught relationship with algorithmic curation. Recent work has also identified that interface design decisions in social media may lead to cognitive issues, such as attention problems and decreased well-being~\cite{lukoff2021design}. As users with mental illness disproportionately experience harms from current social media curation systems, centering their experiences can reveal mechanisms of harm that affect all users and identify platform design opportunities that benefit everyone~\cite{costanza2020design}.

In this work, we focus on how users with mental illness interpret their interactions with social media platforms and algorithmic curation, and how system design affects their mental well-being. To achieve this, we conducted seven design workshops focused on two tasks designed to explore our interest in social media platforms: first, an activity to discover their current experiences with platforms, and second, a prototyping activity to design their ideal platform to support their mental well-being. These workshops were conducted with 21 participants, all of whom used social media platforms with algorithmic curation and identified as having anxiety, depression, or PTSD. We draw inspiration from design justice~\cite{costanza2018design} and designing from the margins~\cite{erete2018intersectional} to inform our approach. Designing with people with mental illness is vital, as conventional design processes often fail to account for their distinct needs~\cite{costanza2020design}. Design justice emphasizes how technological systems can perpetuate systemic inequities and underscores the importance of inclusive design processes that fairly distribute benefits and harms, ultimately leading to better designs for everyone~\cite{costanza2020design,costanza2018design}.

Our results highlight the cause-and-effect explanations, or folk theories, that our participants created to explain their experiences interacting with algorithmic curation. Combining these folk theories from our participants with users' experiences from prior work, we document a pervasive breakdown in the design of social media algorithmic curation that we conceptualize as \textit{entanglement}. We propose \textit{entanglement} as a framework in which participants cannot make clear connections between the actions they take when interacting with social media, the changes they observe in their curated feeds, and affective outcomes on their well-being. Entanglement extends Norman's action cycle~\cite{norman2013design} by introducing six types of errors: three related to the gulfs of execution (``Hiding'', ``Guessing'', ``Flattening'') and three to the gulf of evaluation (``Dangling'', ``Overloading'', ``Disempowering''). Using the lens of entanglement, we identified novel interaction pathways that contributed to many of these breakdowns and led to negative affective consequences for our users. Our participants actively tried to combat entanglement through their design prototypes with three strategies: 1) contextualizing engagement to preserve users' emotional intentions, 2) consumption control for healthier interactions, and 3) reclaiming control through explicit user input.

We introduce a novel framework, entanglement, for explaining the problems users with mental illness encounter when interacting with current social media platforms and resulting harms to mental well-being. Entanglement extends Norman's action cycle through the idea of ``joined worlds'', which explains how merging recommender systems and social networking systems in social media platforms creates conflicts in users' mental models and interaction experiences. Finally, we propose design implications as starting points for combating entanglement through: 1) creating avenues for users to exert control through communicating their intentions and contextual information relevant to their interactions, and 2) shifting algorithmic optimization from accuracy to focus on contextual and emotional aspects of content and users.

\section{Related Work}
\subsection{Social Media and Mental Well-being}
Research consistently shows the importance of social media for people with mental illness: it provides a source of accessible 
information~\cite{milton2023see,pretorius2022mental} and community~\cite{de2017social,naslund2020social,sit2022youth} for many. Social media makes it easy to connect with others about mental health through communities~\cite{de2014mental}, hashtags~\cite{chancellor2016thyghgapp,feuston_beyond_2018}, and social media feeds~\cite{milton2023see}. Previous work has found that people with mental illness will look for social support from others with similar diagnoses~\cite{naslund_future_2016,basch2022deconstructing}.

However, this same research recognizes a trade-off in the risks versus benefits of social media for people with mental illness~\cite{naslund2019risks,karim2020social}. Social media can propagate
unreliable information and malicious misinformation~\cite{lau2012social,basch2022deconstructing}, as well as dangerous content that encourages harmful mental health behaviors and outcomes~\cite{chancellor2016thyghgapp}. More recently, people have begun to use social media for self-diagnosis, creating mixed consequences to users and healthcare providers through pathologizing of normal behavioral and normalizing mental health disorders to combat stigma~\cite{corzine2024inside,hasan2023normalizing}. Given the potential influence of social media on people's mental well-being, platforms have created interventions and policies to try to mitigate potential harms, such as screen timers, content moderation of dangerous content, and nudges to get off the platform. However, these policies can harm users by restricting access to content and communities that provide positive social support~\cite{feuston2019everyday,feuston2020conformity} and silence users from marginalized identities speaking on their lived experiences~\cite{lookingbill2024there}. 

Past research has established that social media is essential to users with mental illness, yet can also have complicated effects on users' mental well-being. In this work, we use the context of social media and mental health as a case study of algorithmic curation to examine the consequences to users and ways to design these systems to support them.

\subsection{Algorithmic Curation in Social Media}
Next, we provide an overview of the research on the effects of recommender systems/algorithmic curation on social media and users. Recommender systems and their impacts on users are a popular area of research~\cite{konstan2012recommender,knijnenburg2012explaining}. In recent years, recommender systems have been used on social media feeds to provide content matching users' tastes~\cite{berman2020curation}, which is often called algorithmic curation~\cite{gillespie2014relevance}. Algorithmic curation of social media can improve connectivity~\cite{berman2020curation} and the delight of discovering new content~\cite{milton2023see}. However, other research finds that algorithmic curation polarization and filter bubbles~\cite{berman2020curation,cho2020search,celis2019controlling}, as well as echo chambers and rabbit holes~\cite{brown2022echo,o2015down,ledwich2019algorithmic}.

The study of algorithmic curation in social media has evolved from initial explorations of user perceptions to more nuanced examinations of user experiences as algorithmic curation has become more pervasive. Early work in this area found that users were not always aware that algorithmic curation was occurring~\cite{eslami2015always}. They reasoned about algorithmic curation in nuanced ways~\cite{eslami2016first} and developed elaborate and nuanced folk theories of algorithms~\cite{devito2017algorithms,eslami2016first}. Follow-up work highlighted a critical gap between user perceptions and the actual functioning of social media feeds driven by algorithmic curation~\cite{bucher2019algorithmic,fletcher2019generalised,swart2021experiencing}. More recently, research on algorithmic curation has examined how people engage with algorithms, finding that users struggled with sense-making and domestication of the algorithm that caused issues with their digital selves~\cite{swart2021experiencing,simpson2022tame}. The effects of algorithms on people were bidirectional, with studies showing both increases to mental health symptoms and coping mechanisms and social support from social media use~\cite{gao2020mental,keles2020systematic,milton2023see,milton2024seeking,o2018social}.

\subsection{Effects of Algorithmic Curation on Mental Well-being}
Finally, we address prior work at the intersection of our interests: the impacts of algorithmic curation on mental well-being. This is in light of increased media attention on current social media and mental health~\cite{guardian_paul,cnn_yurkevich, Healthline,ahdhtime}, and the lay hypothesis that feeds harm well-being~\cite{haidt2024anxious}. 

Studies have demonstrated that algorithmic curation can be supportive of mental health. These benefits are wide-ranging, from supporting individuals with information-seeking~\cite{milton2023see,milton2024seeking} to delivering help from professionals~\cite{pretorius2022mental,avella2023tiktok,milton2024seeking} to providing therapeutic value~\cite{joseph2024apps,avella2023tiktok}. Moreover, studies suggest that algorithmic curation can help in mental health discourse~\cite{feuston_beyond_2018,lookingbill2024there} and exposure to people's lived experiences of mental illness~\cite{milton2023see,milton2024seeking,feuston_beyond_2018}. This work indicates that algorithmic curation can serve as a tool for literacy and education, social support, and good advice for mental health. 

However, much of this same work also points to potential risks of algorithmic curation for mental health. This concern is underscored by ongoing public dialogue about the consequences of social media on mental health, especially in adolescents~\cite{haidt2024anxious,naslund2020social,valkenburg2022social}. Studies show that social media can expose users to harmful content, including dangerous advice about eating disorders~\cite{chancellor2016thyghgapp} and methods for suicide~\cite{milton2022users}. Problems with algorithmic curation and frustrations about control can lead to strong reactions by users~\cite{smith2022recommender,feuston2019everyday} and the perpetuation of social inequalities~\cite{swart2021experiencing}. A notable concern is that users report that algorithmic curation can trigger or exacerbate ill-being~\cite{milton2023see,milton2024seeking} or risky behaviors~\cite{dyson2016systematic,feuston2019everyday}. Close to our work, ~\citet{milton2024seeking} identifies frustrating experiences with user interfaces and algorithmic curation, where their participants likened feeds to ``dopamine slot machines''. In this case, participants blamed specific interface characteristics for their ill-being and believed they had no control over the feed.

In sum, prior work emphasizes through folk theories~\cite{devito2017algorithms,eslami2016first} and user experiences~\cite{milton2023see,milton2024seeking,simpson2022tame} that algorithmic feeds may be to blame for people's distressing experiences. We build on the call from work close to ours~\cite{milton2023see} by incorporating and consulting people with mental illness on the design of such systems~\cite{ramazanoglu2002truth,kornfield2022meeting,bardzell2010feminist}. Taking such a standpoint is crucial because individuals with mental illness may disproportionally be affected by the harms of algorithmic curation~\cite{andalibi2020human,feuston2019everyday,feuston2020conformity}.

\section{Methods}
We conducted seven design workshops with 21 participants who identified as having a mental illness and used social media. During these workshops, participants completed two tasks: one aimed at examining their experiences with algorithmic curation and its impact on their well-being, and a design task where they imagined ways to improve social media platforms to support their mental well-being. The study was approved by the Institutional Review Board at the University of Minnesota (STUDY00019753). Here, we outline information about our participants, workshops, analysis methods, and positionality statement.

\captionsetup[table]{belowskip=0pt,aboveskip=0pt}
\begin{table}[h]
\centering
\resizebox{\linewidth}{!}{%
\begin{tabular}{|r|l|r|l|}
\hline
\multicolumn{1}{|l|}{\textbf{Demographic Variables}} &
  \textbf{N} &
  \multicolumn{1}{l|}{\textbf{Demographic Variables}} &
  \textbf{N} \\ \hline
\rowcolor[HTML]{C0C0C0} 
\textbf{Age} & & \textbf{Ethnicity} & \\
18-20 & 5 & African American & 2 \\
21-25 & 8 & Arab American & 1 \\
26-35 & 6 & Asian & 1 \\
36-45 & 1 & Indian & 1 \\
46-55 & 1 & Latino & 1 \\
\cellcolor[HTML]{C0C0C0}{\bf Gender} & \cellcolor[HTML]{C0C0C0} & Middle Eastern & 1 \\
Female & 13 & Turkish & 1 \\
Male & 5 & White & 17 \\
Non-binary & 1 & \cellcolor[HTML]{C0C0C0}{\bf Education} & \cellcolor[HTML]{C0C0C0} \\
Queer & 1 & Bachelor's degree & 8 \\
Transfemme & 1 & Doctorate degree & 2 \\
\cellcolor[HTML]{C0C0C0}{\bf Sexuality} & \cellcolor[HTML]{C0C0C0} & Master's degree & 3 \\
Asexual & 1 & Some college, no degree & 8 \\
Bisexual & 7 & \cellcolor[HTML]{C0C0C0}{\bf Income (USD)} & \cellcolor[HTML]{C0C0C0} \\
Heterosexual & 5 & 1 - 9,999 & 10 \\
Hetero-flexible & 1 & 10,000 - 24,999 & 4 \\
Lesbian & 2 & 25,000 - 49,999 & 4 \\
Pansexual & 1 & 50,000 - 74,999 & 1 \\
Queer & 3 & 75,000 - 99,999 & 1 \\
Questioning & 1 & 150,000 and greater & 1 \\ \hline
\end{tabular}%
}
\caption{Aggregated Demographic Information of Participants. Participants self-identified on intake; some minor standardization was applied (\eg, ``women'' and ``females'' collapsed). Demographic counts are not mutually exclusive and may not sum to the total sample.}
\label{tab:demo}
\Description{A table of the aggregated demographic information for participants that includes age, gender, education, sexuality, ethnicity, and income.}
\end{table}

\subsection{Participants}
We posted recruiting messages on social platforms that used algorithmic curation, primarily X/Twitter, Instagram, and Reddit. We also sent recruitment emails via mailing lists at the author's institution and contacted past research participants who consented to be invited to future research studies. Recruiting messages included a link to a Qualtrics form that verified people's eligibility, recorded their consent to participate, and collected demographic and social media usage information. Respondents were eligible for participation if they were 18 years or older, had been diagnosed (or self-diagnosed) with depression, anxiety, or PTSD, and engaged with social media platforms that provided recommended content (\eg, TikTok, Instagram, YouTube). Following design justice principles, we centered people with mental illness in our study, as those most affected by algorithmic curation's impact on mental well-being, providing them with essential knowledge for redesigning these systems~\cite{hekman1997truth,costanza2018design}. We chose depression, anxiety, and PTSD as our focus conditions based on their high prevalence in the United States~\cite{NIMH_2023}, enabling our findings to reflect the experiences of a significant portion of social media users with mental illness. The survey also included a modified University of California, San Diego Brief Assessment of Capacity to Consent (UBACC)~\cite{jeste2007new} to assess participants' ability to consent to our study, per our IRB's requirements for including people with mental illness.

The first two authors conducted workshops from March to July 2024 until theoretical saturation was reached~\cite{charmaz2012qualitative}. 21 respondents met our eligibility requirements, scored above 15 on the UBACC (which demonstrated capacity to consent), consented to our study, and ultimately signed up for a workshop. Participants received their choice of a \$40 gift card to Target or Amazon for their participation.

Participants were aged 18 to 55 (M=25.5), and were primarily white (N=17) and female (N=13). Our participants also, on average, held a post-secondary degree. Table~\ref{tab:demo} overviews the participants' demographics. We did not collect information about participants' diagnoses of depression, anxiety, or PTSD for eligibility. Instead, we requested self-identification of these conditions, rather than formal assessment, due to well-known socioeconomic and stigma barriers to accessing diagnostic mental health care~\cite{coombs2021barriers} and to align with the intentions of standpoint theory~\cite{bardzell2010feminist}. Participants reported on which platforms they used, with Instagram, YouTube, and TikTok each mentioned by over half of the pool. All these platforms feature videos as a significant form of content. Most participants reported using one of these platforms at least daily, with only two indicating a frequency of weekly or less.

\subsection{Workshops}
We used design workshop methodologies~\cite{orngreen2017workshops} to capture collective perspectives and foster intervention in designs through participants' shared experiences. Given the sensitive nature of mental well-being, we carefully designed our workshops to be supportive environments. Prior research shows that workshops with people with mental illness can increase social inclusion and positively impact well-being~\cite{slattery2020participation,saavedra2018recovery} and can be productive for creative brainstorming and generative engagement~\cite{white2022workshop}. To mitigate potential risks and make participants feel more comfortable, we kept workshop sizes small (2 to 4 participants) and only took notes on participants' discussions in paraphrases and captured pictures of the resulting artifacts. Seven workshops were held: four in person and three remotely via Zoom to improve recruitment. 

Participants who met the eligibility requirements were invited to one design workshop based on their availability. Each workshop took about two hours. Workshops began with introductions, setting ground rules, and icebreakers to help participants feel more at ease with the researchers and with one another. Most of the workshop was dedicated to the two study activities: one focused on exploration and discovery, with the other on prototyping~\cite{spinuzzi2005methodology}. Researchers provided prompts for each activity and asked follow-up questions to facilitate discussion. Throughout the workshop, participants were reminded to reflect on their mental well-being experiences while using their preferred algorithmically curated social media platforms.

\begin{figure*}
     \centering
     \begin{subfigure}[b]{0.53\textwidth}
         \centering
         \includegraphics[width=\textwidth]{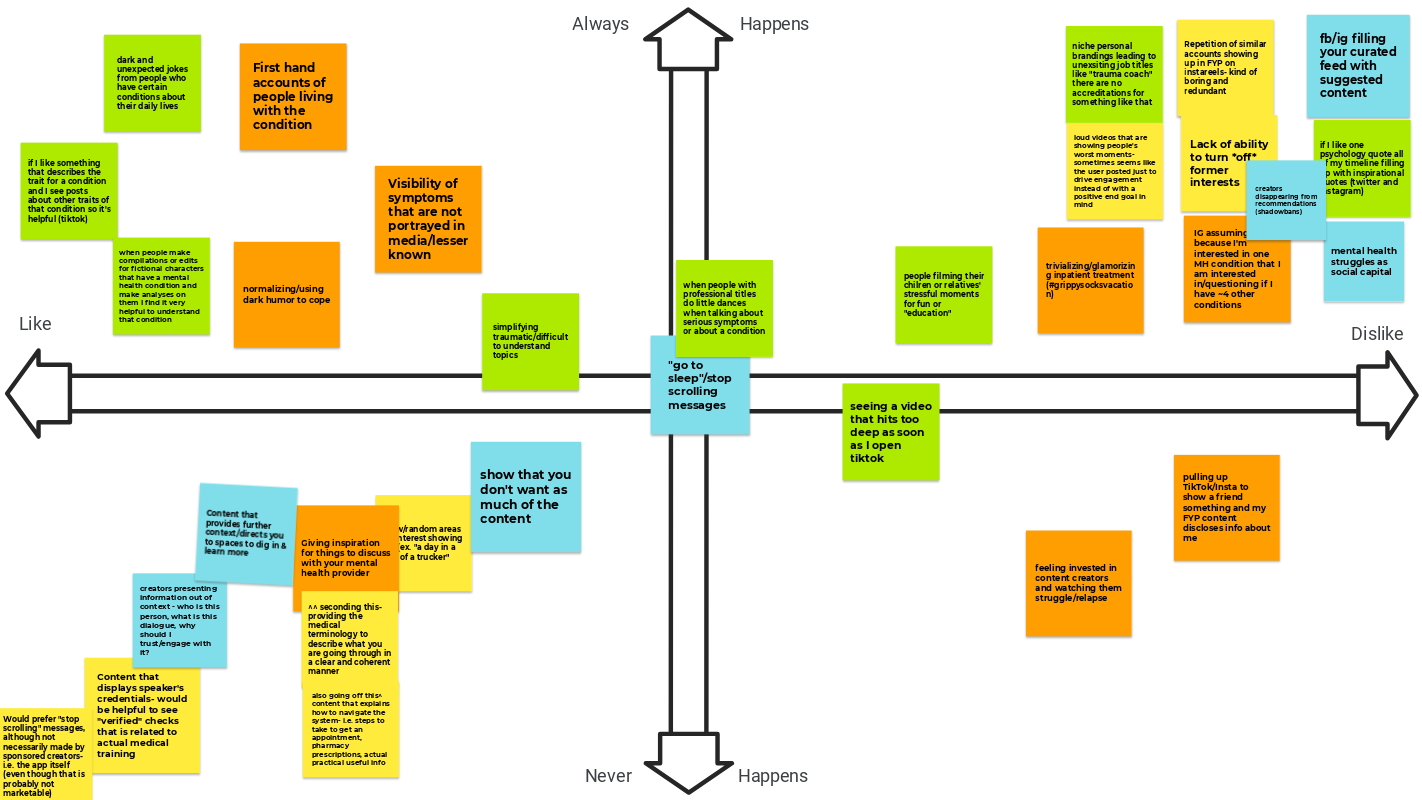}
         \caption{Activity 1: Exploration and Discovery using a Google Jamboard for a remote session.}
         \label{fig:task1}
         \Description{An example of the results from the first task. Colored notes are arranged along a set of horizontal and vertical axes. Each color identifies a single participant who contributed concepts during the session.}
     \end{subfigure}
     \hfill
     \begin{subfigure}[b]{0.42\textwidth}
         \centering
         \includegraphics[width=\textwidth]{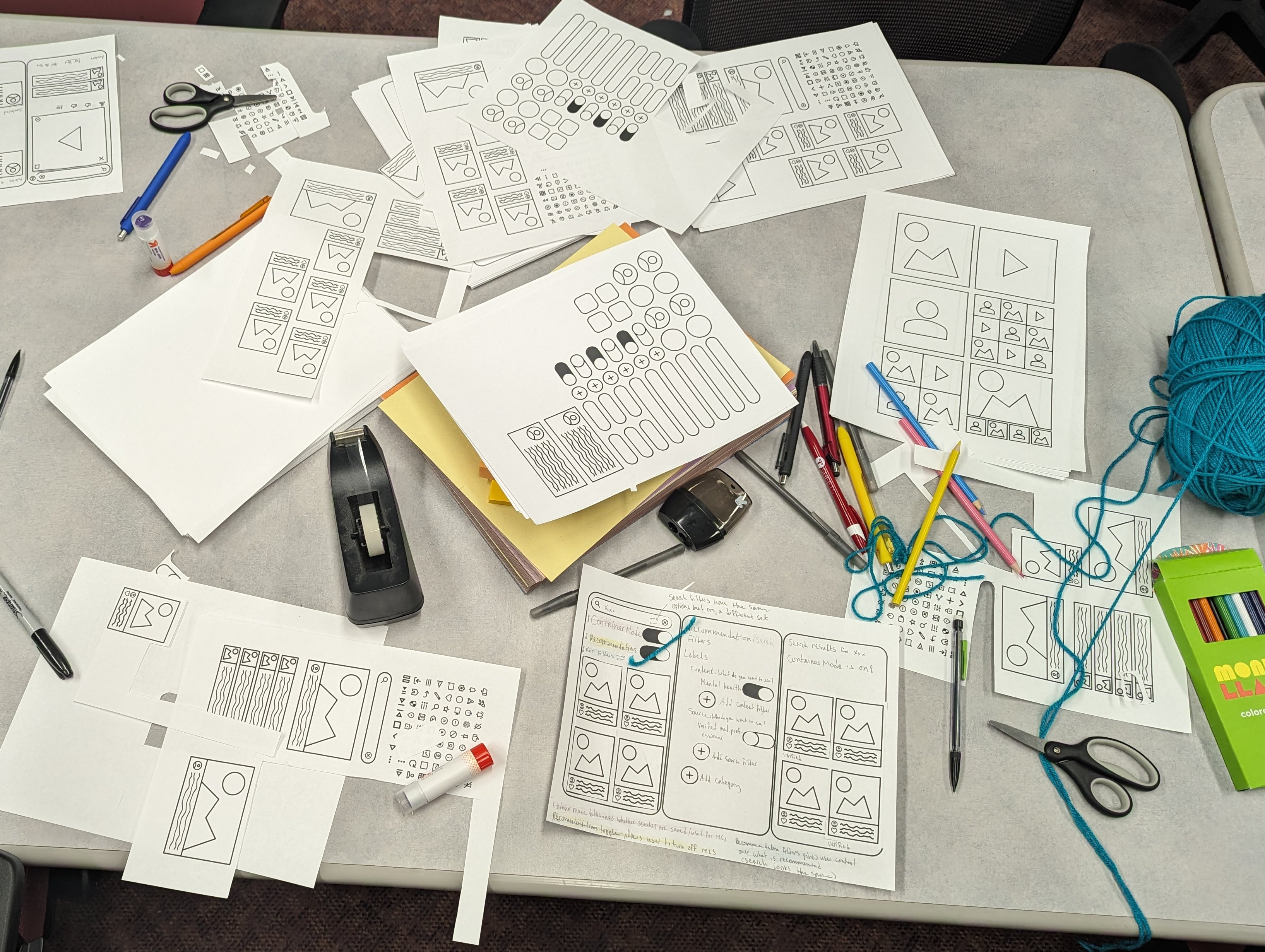}
         \caption{Activity 2: Prototyping from an in-person session.}
         \label{fig:task2}
         \Description{An image of the workspace after the second task. Distributed along the surface are various office and crafting supplies, with paper sheets of common material design components. Examples of some participant designs can be seen with the materials they used.}
     \end{subfigure}
    \caption{Examples of workshop's task outcomes.}
    \label{fig:tasks}
    \Description{This figure contains an image for each task. On the left is the example for the exploration and discovery phase with various comments and ideas arranged in a subjective 2D space. On the right is the workspace after participants constructed their own solutions to the issues they discovered in the previous phase.}
\end{figure*}

\subsubsection{Activity 1: Exploration and Discovery}
For the first activity, participants were asked to discuss their interactions with algorithmically curated social media platforms and to identify the impacts on their mental health and well-being. This activity was modeled after the critical thinking exercise referred to as ``always, sometimes, never''. We provided a shared board for participants to visually orient their experiences along two axes: how often interactions occurred (from ``always'' to ``never'') and how participants felt about the effect those experiences had on their mental well-being (from ``like'' to ``dislike''). Figure \ref{fig:task1} shows an example of this activity. We specifically prompted participants to "think about the interactions that you have had (or wish you had) with your preferred algorithmically curated social media feed," then asked them to write down the interactions and place them on the board. Participants were encouraged to discuss their experiences with one another if they felt comfortable, enabling them to identify common trends and insights. This task allowed participants to explore and recall their interactions with platforms.

During the activity, the facilitating researchers would ask questions to clarify understanding and connection to algorithmic curation when necessary to keep the discussion focused. Researchers further asked follow-up questions at the end of the activity about the interactions provided and where they were placed on the board to promote discussion and gain further insights into participants' experiences and their effects on mental well-being.

\subsubsection{Activity 2: Prototyping}
In the second activity, participants were asked to design an ideal algorithmically-curated social media platform, focusing on how they envisioned the system supporting their mental well-being. The prompt given to participants stated, ``Building on the experiences you talked about in the last activity, design your ideal algorithmically curated social media platform, focusing on how you want it to support your mental well-being.'' The goal was to have participants generate a prototype based on the individual and collective experiences they had identified in the previous task, as well as their ideas for improving the platforms. Toward the end of the sessions, participants shared their prototypes with the group and discussed in a design show-and-tell~\cite{akama2007show}. Throughout the activity, the researchers would ask questions, when relevant, about participants' discussions to understand their motivations for design ideas, connections to interactions from the previous activity, and how they connected to algorithmic curation and promoting mental well-being. Given our focus on design based on participants' needs, researchers refrained from providing any input on the designs. Researchers did ask about design aspects and connections to existing platform features at the end of the activity.

In-person sessions provided office and crafting supplies (see Figure~\ref{fig:task2}) to facilitate the creation of participant prototypes. Remote session participants had a dedicated space on a Google Jamboard slide where they could draw, type, and paste content to help them convey their designs. All participants were encouraged to design in a way that they felt most capable of communicating their ideas (\eg, drawing, writing). Participants were also free to continue discussing their experiences while they worked, and many did.

\subsection{Thematic Analysis}
We used inductive thematic analysis for our qualitative results~\cite{clarke2015thematic,neuendorf2018content}. Themes emerged from analyzing the researchers' notes and the artifacts collected from the two tasks. Between workshop sessions, researchers conducted initial open coding and code collating. The first two authors conducted a secondary analysis phase after all workshops to identify underlying conceptual threads connecting emerging themes. During this secondary analysis, the authors noted a prevalent trend in users' interactions with social media, which led to the development of the entanglement framework (see Section~\ref{sub:entangle}) to reinterpret the emergent theme. Additional discussions between all authors confirmed the resulting findings. All quotes are either from field notes or the artifacts produced by participants. 

\subsection{Positionality Statement}
Most authors of this study identify as active or peripheral members of the mental health research community~\cite {adler1987membership} and are active participants on social media platforms. We believe that including community members in the research team leads to more thoughtful research practices and more open dialogue among participants in at-risk populations. However, including member researchers does not inherently protect us from introducing bias. 

The drive to create thoughtful research practices led us to adopt a design justice approach~\cite{costanza2018design,bardzell2010feminist} in our work by centering on participants' needs (\eg, small workshops, acknowledging self-diagnosis), which we believe promoted better inclusion for people with mental illness. All authors were involved in the study's design, refinement, and presentation, while the first two were responsible for implementing and managing the workshops.

\section{Findings}
Our findings show that users experience emotional issues when interacting with their algorithmically curated feeds through \textit{entangled} interactions. In Section~\ref{sub:folk}, we present participants' explanations for why they felt they were having issues in their experiences, described through cause-and-effect folk theories. Some of the findings in Section~\ref{sub:folk} echo prior work; however, these folk theories provide evidence that participants and others face fundamental issues in the design of social media algorithmic curation. In Section~\ref{sub:entangle}, we introduce the framework of \textit{entanglement}, where users cannot clearly connect their actions to observed system responses, to explain these issues. We describe six forms of entanglement derived from participants' experiences. Finally, Section~\ref{sub:design} presents participants' proposed design as solutions to address entanglement, focusing on contextualizing engagement and returning explicit feed control to support mental well-being.

\subsection{Folk Theories of Algorithmic Experiences}
\label{sub:folk}
During our workshops, participants expressed frustration with algorithmically curated feeds and explained the emotional experiences they had from algorithmic interactions. Following seminal work by~\citet{devito2017algorithms} and~\citet{eslami2016first}, we describe their insights as a set of {\it folk theories} used to describe feed behaviors, which made their experiences emotionally taxing. Folk theories are ``intuitive, informal theories that individuals develop to explain the outcomes, effects, or consequences of technological systems''~\cite{devito2017algorithms}. Through relying on intuitions about the algorithm's underlying mechanisms and intentions, participants made explicit cause-and-effect relationships about algorithmic behavior and emotional outcomes.

\subsubsection{Theory of Emotional Mismatch}
Participants theorized that algorithms lack an understanding of ``emotional weight'' (P13), particularly in content. Unpredictable emotional shifts in feeds create a misalignment with users' emotional states. Participants described feeds as emotionally volatile, reasoning that they do not account for this context. When content is emotionally unpredictable, it leaves users unprepared for emotionally burdensome content.

For our participants, social media content is laden with emotional variation. P11 described feed content emotionally ranging from ``happy moments and people being happy’’ to ``messy breakups’’. Participants believed feed algorithms disregarded this emotional context, treating all content as emotionally equal. This situation created what P20 described as ``recommendations don't always match the mood at the time'' which emotionally burdened them. Similar observations were reported by~\citet{milton2023see}, who described feeds as ``runaway trains'' that deliberately push negative emotions.

Our participants mainly focused on the emotional misalignment between their feeds and their own emotional needs. P18 often experiences being shown a ``heavy video'' on mental health topics when first opening an app, affecting their ``whole mood'' for the day. Having no control over when emotionally heavy content appeared was compounded by difficulties in ``turning off'' (P18) content, as it was ``easy to start getting recommendations for specific topics'' (P19) but not stop them. The sudden shift in emotion, or what our participants routinely called ``mood swings'', occurs without warning, leaving users without ways to prepare their mental state.

Emotional misalignment did not always have a negative outcome; mismatches sometimes led to serendipitous discoveries and positive relational outcomes~\cite{cullen2025not}. Participants did not want to eliminate emotional mismatches, as content could provide validation through ``other people's hyper-specific experiences'' (P10) or examples of disability and personal struggles (P20), and be valuable when their emotional state aligned (P18). ~\citet{lee2022algorithmic} and ~\citet{simpson2022tame} report similar feelings of support and validation from feeds that represent aspects of the self. Emotional volatility in feeds could also facilitate connections and support, helping participants to ``connect with friends and family over videos and popular topics'' (P15) through pebbling~\cite{cullen2025not}. Participants believed the same algorithmic mechanism both imposed emotional burdens and provided support. This contrast highlights their desire for control over emotionally weighted content.

\subsubsection{Theory of Algorithmic Leakage}
Participants saw their algorithmic feeds as unintentional disclosure agents, ``leaking'' personal information through recommended content. In this theory, feeds function as mirrors~\cite{lee2022algorithmic,cullen2025not}, reflecting users' identities through content. As feed content may be visible in shared or unsafe places, participants felt their private information, especially related to mental well-being and emotions, was exposed, prompting hypervigilance.

Passive mental health disclosures were a primary concern, showcasing ongoing stigma and fear of judgment around mental well-being. Participants noted that recommendations reveal prior interactions, with P04 saying, ``everyone knows that you get recommendations based on what you have already seen'', and P17 adding ``I feel like I will be judged for having mental health content on my feed.'' While prior work has shown that feeds reflect users' identities~\cite{lee2022algorithmic,cullen2025not} and can even influence their self-concept~\cite{cotter2022fyp,karizat2021algorithmic}, our participants emphasized the risks of passive mental health disclosure in shared spaces. Concerns about unintentional disclosure increased anxiety and reinforced the sense that feeds exposed more than users intended.

Participants' concerns extended beyond feeds, as they perceived that platforms share data with other platforms, increasing disclosure risks. Participants gave examples of ``searching for some [mental health] symptoms, and it seems to have crossed over into my feed'' (P09) or ``cookies on one site can travel to other sites and share the information'' (P04), creating a need for them to be cognizant of how their interactions off-platform bleed into their algorithmic feeds. Together, these beliefs show that even when users are hypervigilant with their feeds, their interactions, particularly with mental health content, may be used in unanticipated ways, increasing the potential for harmful disclosure across platforms.

\subsubsection{Theories of Algorithmic Exploitation}
Our participants described two folk theories about how algorithmic feeds exploit their interactions. They believed these malicious intentions were deliberately done to affect their well-being. Both theories illustrate how participants attempted to make sense of their feed behaviors, which left them feeling distrustful and frustrated.

\textbf{Theory of Corporate Intentions.} Participants believed companies made deliberate choices that affected their mental well-being, evidenced by engagement-driven algorithmic design. Participants felt that ``[companies] know their algorithm is addictive'' (P10). They reasoned that corporations implemented algorithms to exploit people's interactions and engagement to make money by keeping them on the platform (P05). Our participants attributed strong intentionality to an obscured process of algorithmic curation to understand their experiences, similar to prior issues with TikTok ``techlash''~\cite{simpson2022tame} and amplifies the ``Platform Directed'' operationalization seen in folk theories about Twitter's algorithm~\cite{devito2017algorithms}.

The belief that the algorithm was intentionally pushing engagement influenced how participants viewed platform-led mental well-being interventions (P10), like ``stop scrolling'' messages or ``notifications to take a break.'' As participants believed companies acted with exploitative intent, they viewed these interventions as ``very patronizing because it was from the platform'' (P19), provoking anger rather than self-reflection. However, when similar messages came from creators, participants found them helpful (P18). Participants' intuitions that companies' intentions were exploitative made them distrust platforms' efforts to support users' mental well-being. 

\textbf{Theory of Engagement Abuse.} While the previous folk theory focused on the \textit{why}, participants also created beliefs about \textit{how} the algorithm accomplished corporate goals. Participants theorized that the underlying algorithm used every interaction to promote engagement at the expense of mental well-being. 

This belief came from participants' experiences of ``any engagement causes the recommendations to shift'' (P14). Participants felt feeds purposely ``prioritize attention and engagement over users' well-being'' (P04), reinforced by the sense that ``[companies are] filling your feed with stuff {\it they} want you to see, not stuff {\it you} want to see’’ (P19). Participants' belief that platforms deliberately chose outcomes led to increased feelings of distrust, as these engagement-driven outcomes were perceived as manipulative. This distrust, combined with participants' coping with uncertainty in their interactions, contributed to participants' overall experiences of anxiety and frustration when interacting with the platforms.

\subsection{The Entangled Action Cycle}
\label{sub:entangle}
\begin{figure}[h]
    \centering
    \includegraphics[width=0.7\linewidth]{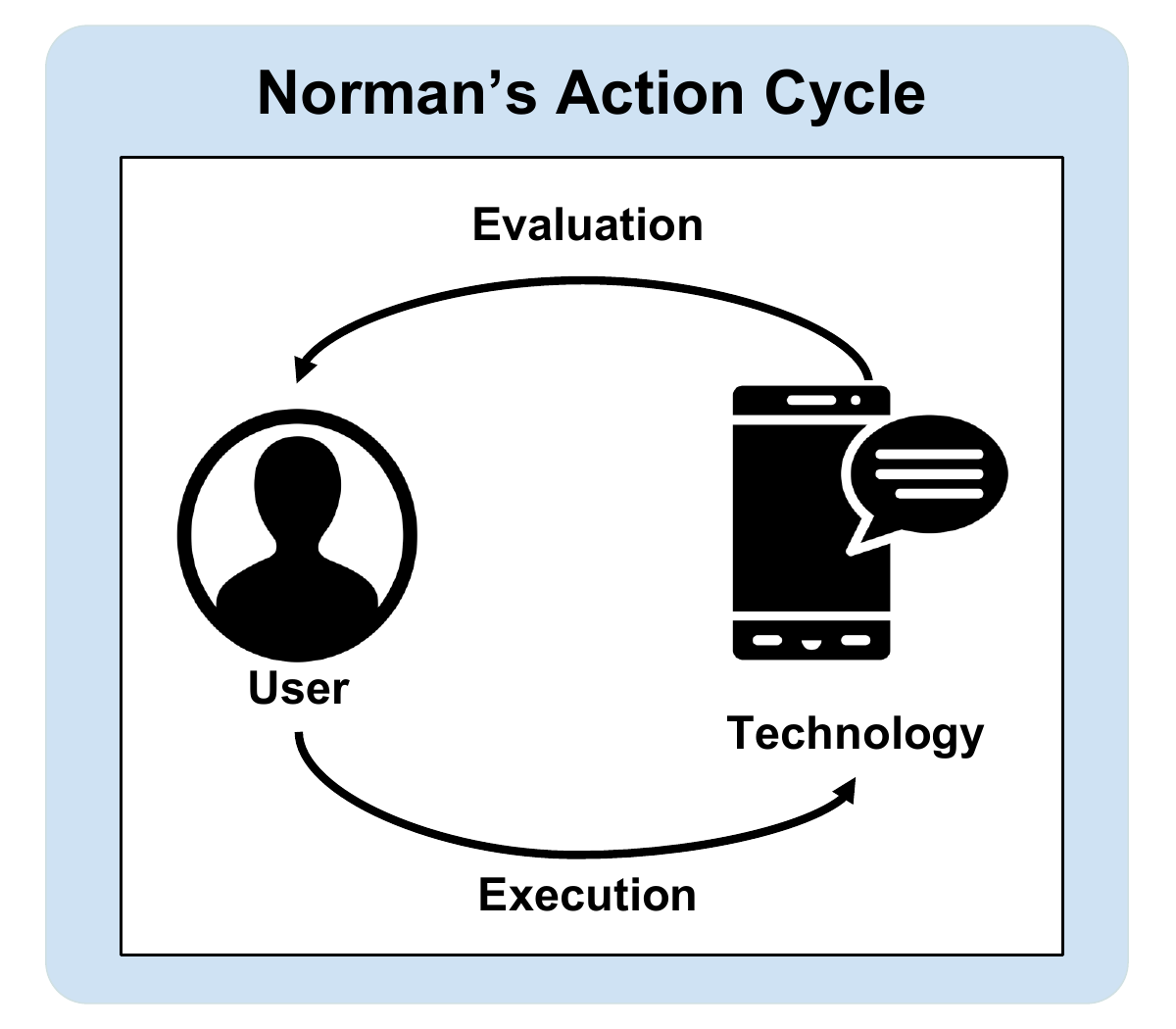}
    \caption{Visual representation of Norman's Action Cycle}
    \Description{An illustration of Norman's action cycle. Contains two icons: a user on the left and a piece of technology on the right, connected by arrows. The arrow from the user to the technology is labeled execution, and the arrow from technology to the user is labeled evaluation.}
    \label{fig:norman}
\end{figure}
From participants' folk theories, we identified patterns in their interactions with social media and the outcomes they saw in their feeds. These patterns reveal a disconnect between participants' perceptions of their actions and changes in their feeds. This disconnect is clearly visible through users with mental illness experiences, who are more attuned to emotional changes and interact with greater emotional intentionality. To understand this disconnect, we developed the framework of \textit{entanglement}, which describes participants' inability to make clear connections between the actions they take on a platform and the resulting system responses they observe. This framework was developed during a secondary analysis phase to reinterpret emergent themes in participants' experiences arising from interactions with social media algorithmic curation. In this section, we describe our framework of entanglement and the emotional outcomes it had on participants.

Our framework builds on Norman's action cycle~\cite{norman2013design}, popularized in the {\it The Design of Everyday Things.} The action cycle is a seminal psychological model that describes how people interact with objects in the world. Norman posits that people ``continually act upon and monitor the world, evaluating its state...[and] executing actions''~\cite{norman2013design}. This approach aligns with our participants' folk theories that describe a cause-and-effect model of their interactions. By situating participants' accounts of their interactions within the action cycle, we can more clearly examine their experiences to identify where these disconnects, or entanglements, occur.

In HCI, this action cycle has four components: user, technology/ artifact, execution, and evaluation~\cite{norman2013design}.  As illustrated in Figure~\ref{fig:norman}, the cycle starts with a user having a goal they wish to accomplish, which they execute through iterations with the relevant technology, and then they evaluate that technology's response to determine if they have achieved their goal. For Norman, good design makes this cycle clear and easy. In contrast, poorly designed objects, such as ``Norman doors'' \cite{norman2013design}, lack clear indications of whether they should be pushed or pulled. Thus, bad design causes users problems when using the underlying object to achieve their goals. Problems that emerge in this cycle are classically described as \textit{gulfs of execution} and \textit{gulfs of evaluation} (or less commonly, slips~\cite{norman1981categorization}). 

Currently, gulfs in the action cycle neglect a critical dimension of participants' experiences: emotions. These gulfs take a mechanical view of users' interactions or possibly contextualize this as user error~\cite{norman1981categorization}. However, our participants noted the affective/emotional aspects to their intentions and experiences with algorithmically curated feeds, which were caused by the interface. Entanglement addresses this gap by centering on emotions in user actions and system responses, characterized through six issues that encompass emotional aspects in both the execution and evaluation phases of the action cycle. Thus, entanglement extends the current action cycle to include important emotional aspects relevant to users when interacting with technological systems.

In the rest of the section, we define and show examples of these six categories:
\begin{itemize}
\item Issues relating to execution: \textbf{Hiding}, \textbf{Guessing}, and \textbf{Flattening}. 
\item Issues relating to evaluation: \textbf{Dangling}, \textbf{Overloading}, and \textbf{Disempowering}.
\end{itemize}

\subsubsection{Entanglement Issues in the Gulf of Execution}

\begin{figure*}[h]
    \centering
    \includegraphics[width=0.85\linewidth]{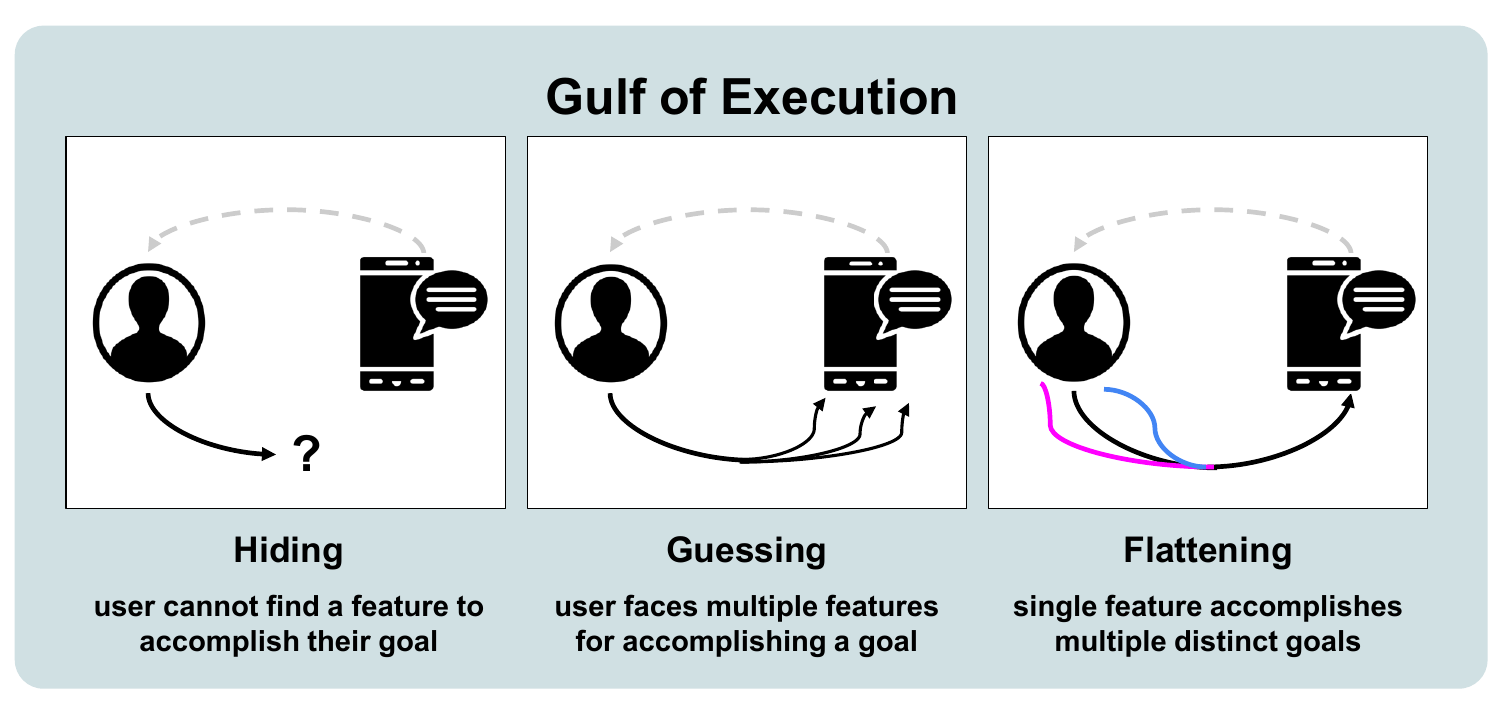}
    \caption{Visual representation of the entanglements in the Gulf of Execution}
    \Description{A 3-panel diagram illustrating the entanglement issues in the Gulf of Execution. Panel 1 (Hiding): Users cannot find features. Panel 2 (Guessing): Users have multiple unclear options. Panel 3 (Flattening): Users have one feature for multiple goals. Each panel shows a user icon connected by different kinds of arrows to an interface icon to represent the different issues.}
    \label{fig:execution}
\end{figure*}

We first consider issues on the execution side of the action cycle, as shown in Figure~\ref{fig:execution}. The issues we outline in this section relate to Norman's \textit{gulf of execution}, the difference between a user's goals and what a system allows them to accomplish~\cite{norman2013design}. They build on this traditional understanding to capture our participants' experiences and emotional responses. 

\textbf{Hiding} occurs when users cannot find a feature to accomplish their goal because the feature is missing or unavailable. Hiding is close to a classic gulf of execution, arising when a user has a single goal but perceives that there are no features to accomplish it. The feature may be hidden, non-existent, or unclear in its functionality. This issue occurred when participants wanted to ``reset’’ their feeds after getting ``stuck in a rabbit hole, and then my feed is trashed’’ (P12), leading to uncomfortable content with no way to correct it. Participants would look for ways to reset, but no options were available. One participant noted their feeds ``somehow went from scientific and gardening videos to bare-feet anti-vaccine content'' (P12). The inability to ``reset'' a feed is particularly an issue for mental well-being and emotions, as participants reported feeling distressed by the stream of content they were viewing. Without recourse to correct their feed, participants experienced intensified emotional impacts. An example of such content was that which romanticizes disabilities and ``belittles experiences'' which they felt ''creates insecurity'' (P14) for them.

Another goal participants could not accomplish was removing content from their feeds. Some content had adverse effects on participants, such as the content mentioned earlier that romanticizes disabilities or content that ``trivializes/glamorizes inpatient treatment'' (P17). Participants wanted to remove such content, not only for themselves but also for others, to mitigate the harm it could cause to people's mental well-being. However, many did not see a way to accomplish this, and participants could not remove it from their feeds. Being unable to remove content left our participants frustrated and continually exposed to content they did not wish to see, which could make them feel uncomfortable or relive distressing experiences. Hiding also aligns with the \textit{theory of corporate intentions}, as hiding or failing to provide mechanisms for users to take actions to influence their feeds can be seen as a deliberate act of forcing engagement through the inability to act at the expense of users' mental well-being.

\textbf{Guessing} happens when a user faces multiple, unclear features for accomplishing a goal, forcing them to experiment to determine which feature will achieve their goal. The most common Guessing scenario related to algorithmic curation was when participants did ``not want to see stuff [specific types of content]'' (P19). Participants identified several features they guessed would help, including ``not interested'', or ``do not recommend''. However, no social media platforms clearly indicated which feature would actually permanently remove content. P01 noted it was ``hard to convey that you're not interested in something'', while P02 felt they had to ``jump through hoops to stop seeing some content''. Guessing is an entanglement focused on the need to engage with multiple features that drive engagement, which aligns with the \textit{theory of engagement abuse} when a purposively ambiguous design forces multiple interactions.

Guessing creates confusion in users' minds, leading them to continue using features with uncertainty to regain control of their feeds. In cases of conflicting features, some participants, such as P12, began blocking creators to avoid viewing specific types of content. Blocking, however, is a creator-focused feature, not a content one, and would exclude an entire creator's content from their timeline. This created tension: participants wanted to see some of a creator's content, but not all of it, and, since they were using blocking as a workaround, they got stuck. One example was a creator whose content spanned ``from ghost video to horror... [it has] similar undertones but the content is extremely different'' (P15). P15 wanted more fine-grained control but couldn't find it. The fact that participants went so far as to use a creator-focused feature like ``blocking'' for a content-focused goal shows how much guessing they had to do to accomplish their goal. Participants noted increased mental burden and emotional exhaustion while trying to remove content already harming their mental well-being.

\textbf{Flattening} is when a single feature accomplishes multiple distinct goals, erasing the nuance of users' contextual needs and what the feature actually claims to do. This issue was prominent in participants' discussions of ``clicks'', ``comments'', and ``watch time'' for short-form videos on platforms like TikTok and Instagram. Participants perceived that the platform inferred these features to mean more than users intended, leading their feeds to become out of control. Our participants confirmed that ``there is no distinction between clicking on something because you like it or hate it. The platforms don't care either way'' (P06). Comments were another area where Flattening came up. Comments accomplish different goals, such as showing support or interest, correcting content when erroneous information is present in a post, or arguing with others. All of these contextual uses of comments are collapsed into a single category: simple positive engagement. Participants noted having to weigh the pros and cons of interacting using such features, particularly in cases where they wished to stop the propagation of stigmatizing mental health content but also did not want their feeds to become inundated with such content, as the sheer volume could cause distress. The \textit{theory of emotional mismatch} is directly connected to flattening, as it reveals the removal of emotional dimensions from actions before algorithm interpretation.

Participants reported that the flattening of the ``watch time'' feature was particularly detrimental to their mental well-being. We note that watch time or basic engagement is a {\it passive} feature on platforms, where users have little control to ``stop'' watching content. Thus, ``watch time'' suffered from the same issue as commenting, as participants noted that platforms ``can't tell the difference between watching something and being satisfied or enraged by it'' (P01). Participants described getting ``sucked into watching the content'' and either ``reveling in self-hatred'' (P01) or finding ``the dopamine from watching short videos can keep you stuck'' (P10). Content such as ``Rage bait'' exemplifies the lack of nuance in watch time and showcases its effects on users as it ``drives engagement instead of with a positive end goal in mind'' (P13). Participants' inability to look away from the content was taken as a measure of their engagement, regardless of whether they liked it or how it affected their mental well-being. These examples show that participants felt a single watch-time metric cannot distinguish between genuine engagement and compulsive viewing, which at times led them to internalize the emotions presented to them. Similarly, the \textit{theory of algorithmic leakage} and the \textit{theory of engagement abuse} are relevant to flattening, as the collapsing of goals to drive engagement can also unintentionally lead to unwanted disclosure.

\begin{figure*}[h]
    \centering
    \includegraphics[width=0.85\linewidth]{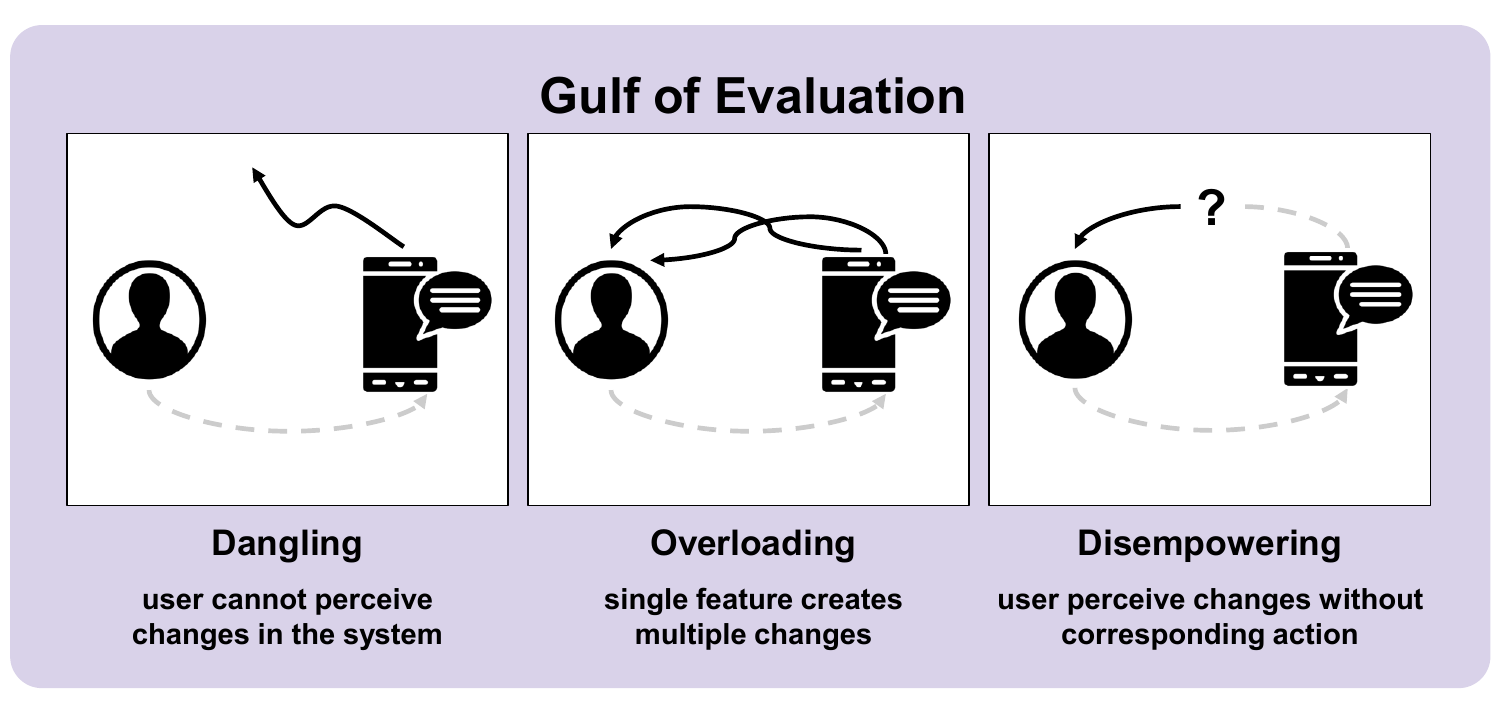}
    \caption{Visual representation of the entanglements in the Gulf of Evaluation}
        \Description{A 3-panel diagram illustrating the entanglement issues in the Gulf of Evaluation. Panel 1 (Dangling): Users cannot perceive system changes. Panel 2 (Overloading): Users perceive multiple system changes. Panel 3 (Disempowering): Users perceive changes without them acting. Each panel shows an interface icon connected by different kinds of arrows to a user to represent the different issues.}
    \label{fig:evaluation}
\end{figure*}

\subsubsection{Entanglement Issues in the Gulf of Evaluation}
\label{entangle:eval}
Next, we move to examining entanglement related to the gulf of evaluation of the action cycle, as seen in Figure~\ref{fig:evaluation}. Recall that gulfs of evaluation are the degree to which the system's state can be interpreted in terms of a user's goals~\cite{norman2013design}. We build on the concept of the \textit{gulf of evaluation} in outlining nuanced issues in current social media.

\textbf{Dangling} occurs when users cannot perceive any changes in the system from their interactions with features. Almost every participant said that features for showing disinterest in content did not work. Some were ``confused about whether clicking on dislike would stop the recommendations'' (P14). Participants described confusion stemming from a perceived disconnect between their emotional intentions, the system's interpretation of their actions, and the subsequent responses. While participants could sometimes find a feature to alter their feed content, the feature seemed to have no effect. Participants initially felt hopeful after discovering these features, but became frustrated and worried when they did not notice an effect. The \textit{theory of emotional mismatch} and the \textit{theory of corporate intentions} support this entanglement, with engagement features designed in ways that are ambiguous and lack emotional consideration, driving emotional misalignment with feed content.

Dangling is exemplified when contrasting``disliking'' and ``liking''. P18 noted the difference between how ``seriously'' feeds responded to their positive engagement (``likes'') versus negative engagement (``dislikes''), if it was available at all. Similarly, participants found ``blocking'' users and channels unhelpful in removing unwanted content. P12 had to ``block them like seven times'' yet ``it still shows me [that content]''. Some felt that the ``algorithm is testing'' them, as content ``will come back'' (P18). Knowing that content could reappear, even if participants could successfully remove it, was noted to increase feelings of anxiety. Participant P07 described having to ``filter your content for yourself, even though the algorithm is supposed to send you content that relates to you''. However, it was unclear ``how an action you're about to take would impact your experience’’ (P04), as participants could not interpret any changes in their feeds. 

\textbf{Overloading} happens when a single feature creates multiple and often conflicting changes in the system that confuse users about the feature's outcome. The most prominent examples of overloading were ``sharing'' and ``liking'' content, which both serve multiple purposes for users. ``Sharing'' allows users to send content to others and build social connections, but participants believe it is also used for positive engagement, affecting both the sender's and recipient's feeds. Participants echoed P17's statement of wishing they ``could share posts but also not send other personal metadata with it'' as ''people seem to get content suggested to them when I share things ... related to topics I engage with'' (P17). The example of sharing illustrates how participants connected their experiences of emotional distress to the unpredictability of overloaded system responses. Participants felt anxious about what would be revealed about them and their mental health status through the metadata sent with the content via sharing. We can observe the \textit{theory of engagement abuse} in overloading, where platforms leverage a single interaction in multiple ways to drive engagement. Interactions that participants did not perceive as connected to engagement clearly support this theory’s connections to overloading.

``Liking'' is also used for both social connections and algorithmic engagement. Participants noticed that ``if I like one psychology quote, all of my timeline fills up with inspirational quotes'' (P18), meaning that ``liking'' also results in more similar content in feeds, regardless of the intention of the user's action. At a creator level, ``the algorithm assumes that if you like one video from a creator, you'll like all their others, regardless of differences in exact content'' (P01). Participants struggled to understand changes to their feeds, as they perceived a mismatch between their intentions and the multiple ways the system used their interactions. Overloading has clear connections to both the \textit{theory of emotional mismatch} and the \textit{theory of algorithmic leakage} through the compression and exchange of emotionally laden information without regard for the emotional or social implications it creates.

\textbf{Disempowering} is when users perceive system changes happening without any corresponding features that give them control over initiating or stopping such changes. Participants experienced this issue through features such as infinite scroll and autoplay, which led to compulsive content viewing. When these features were combined to create a continuous content stream, users felt compelled to engage, regardless of their intentions or the harm it caused to their well-being. Yet the ``watch time'' registered as positive engagement due to ``any engagement is positive and will get you more content’’ (P20). Participants described an emotionally distressing cycle in which they had no agency to escape content that harmed their mental well-being, leaving them frustrated, anxious, and disempowered.

Participants also found that their feeds would drastically change with ``all the progress [personalization] gets reset'' (P10) and showing generic content. Participants did not understand why these changes happened or what caused them. They were left with the heavy mental burden of curating their feeds again and possibly being exposed to content that evoked feelings of distress, rage, or anxiety, which they had previously successfully removed from their feeds. These examples showcase how participants struggled to connect their interactions with system responses, leaving them unable to control or manage their feeds and, in some cases, feeling demoralized and confused. The \textit{theories of algorithmic exploitation} are the foundation of the entanglement of disempowering with its direct connection to concepts of forced and misrepresented engagement to drive compulsive consumption, and thus profits.

\subsection{Mitigating Entanglement Through Design}
\label{sub:design}
Participants worked to resolve entanglements caused by algorithmic curation to protect their mental well-being. Our analysis reveals three major design themes our participants proposed to address entanglement: contextualizing engagement to preserve users' emotional intentions, consumption control for healthier interactions, and reclaiming control through explicit user input. Each high-level design theme is attempting to mitigate entanglement in the gulfs of execution and evaluation that our participants experienced.

\begin{figure*}
    \centering
    \begin{subfigure}[c]{0.35\textwidth}
        \centering
        \includegraphics[width=\textwidth]{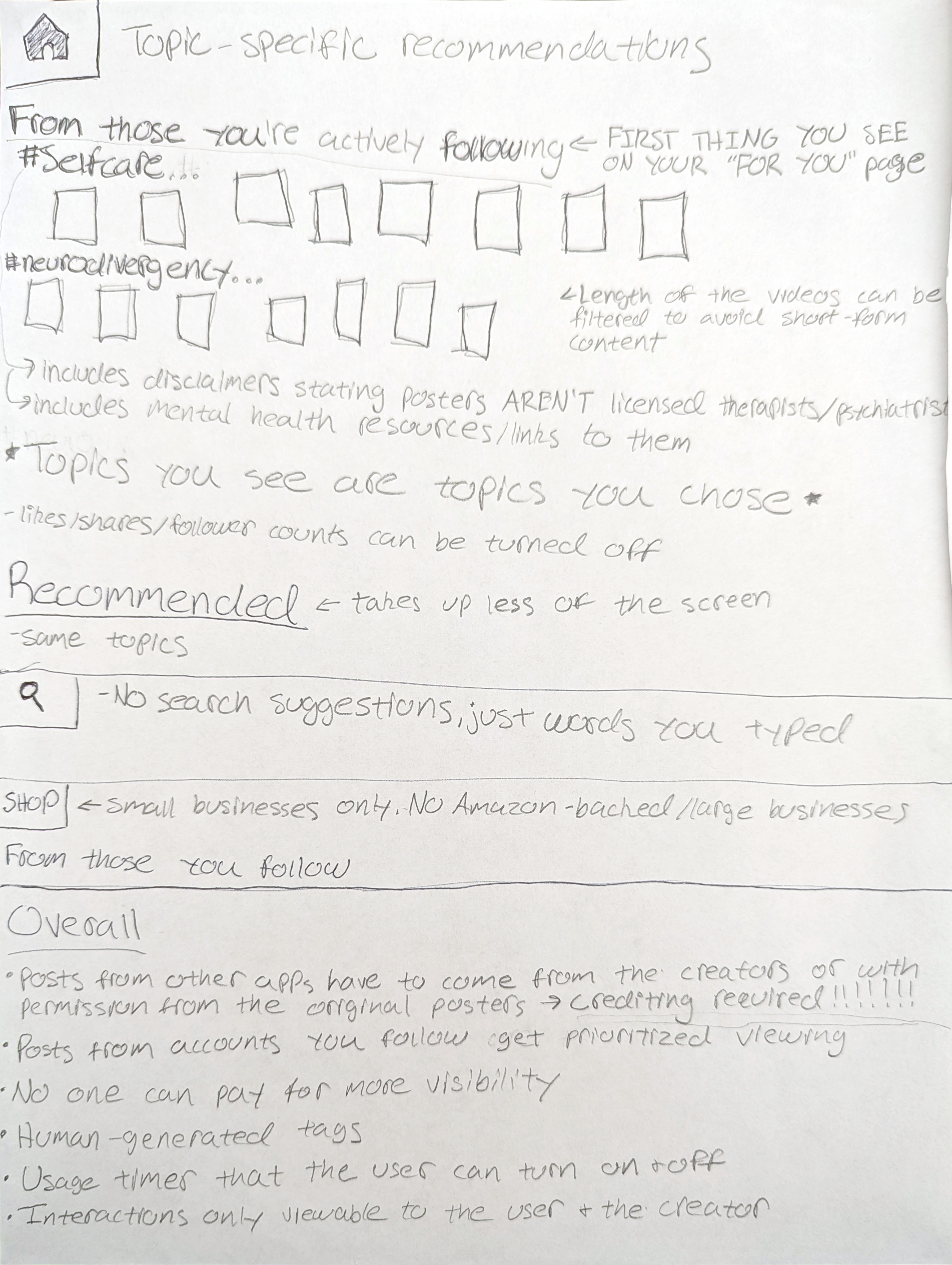}
        \caption{Design ideas from P09.}
        \label{fig:P09}
        \Description{An image of a hand-drawn interface design. The design features 3 sections: topic-specific recommendations, search, and shop.}
    \end{subfigure}
    \begin{subfigure}[c]{0.5\textwidth}
        \centering
        \begin{subfigure}{\textwidth}
            \centering
            \includegraphics[width=\textwidth]{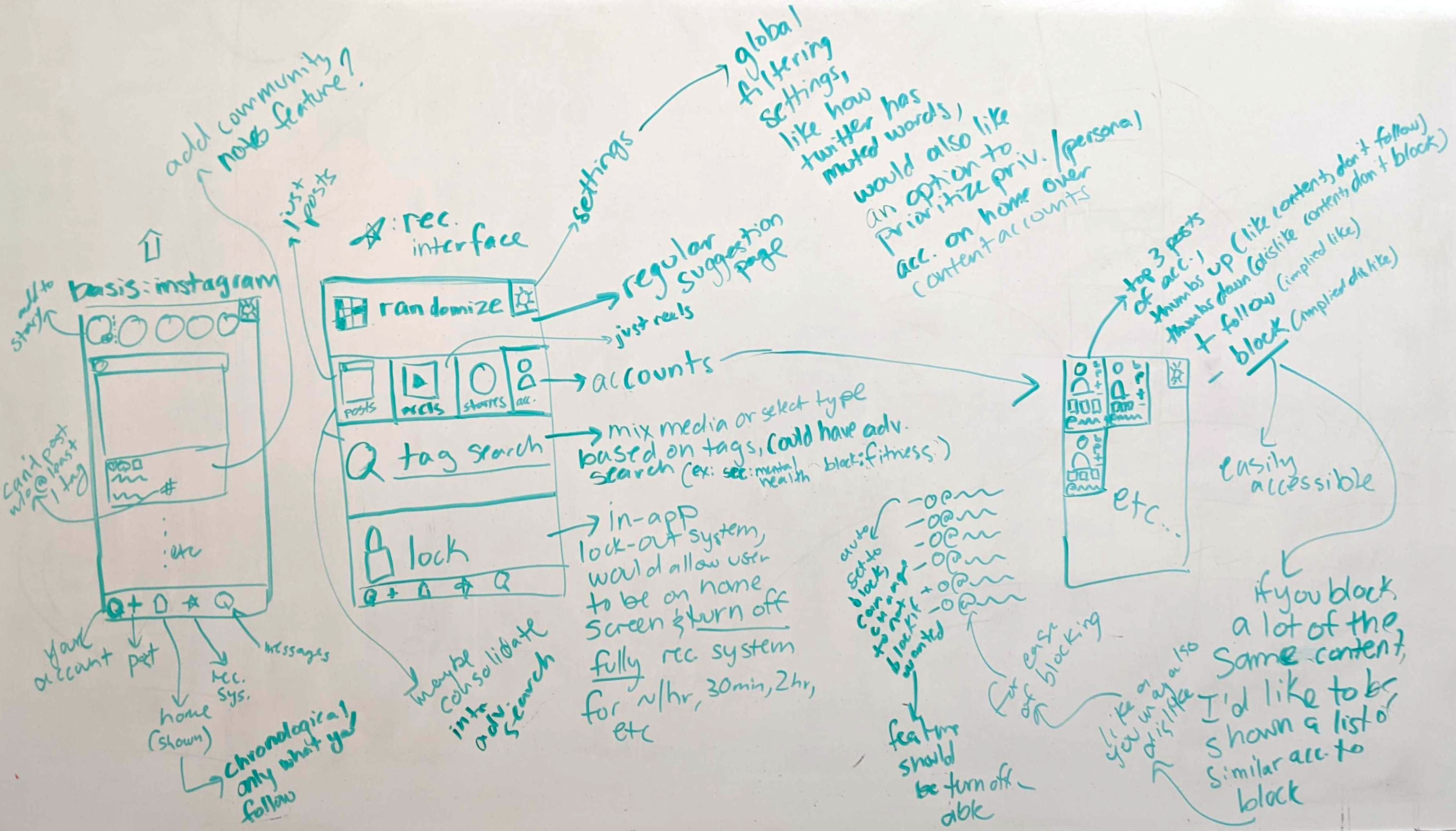}
            \caption{Design ideas from P12.}
            \label{fig:P12}
            \Description{An image of a hand-drawn interface design. The design features 3 interface screens: one for scrolling posts, another with features for controlling the feed through tags, filters, and engagement locks, and the last allows for specific creator and topic following and blocking.}
        \end{subfigure}
        \begin{subfigure}{\textwidth}
            \centering
            \includegraphics[width=\textwidth]{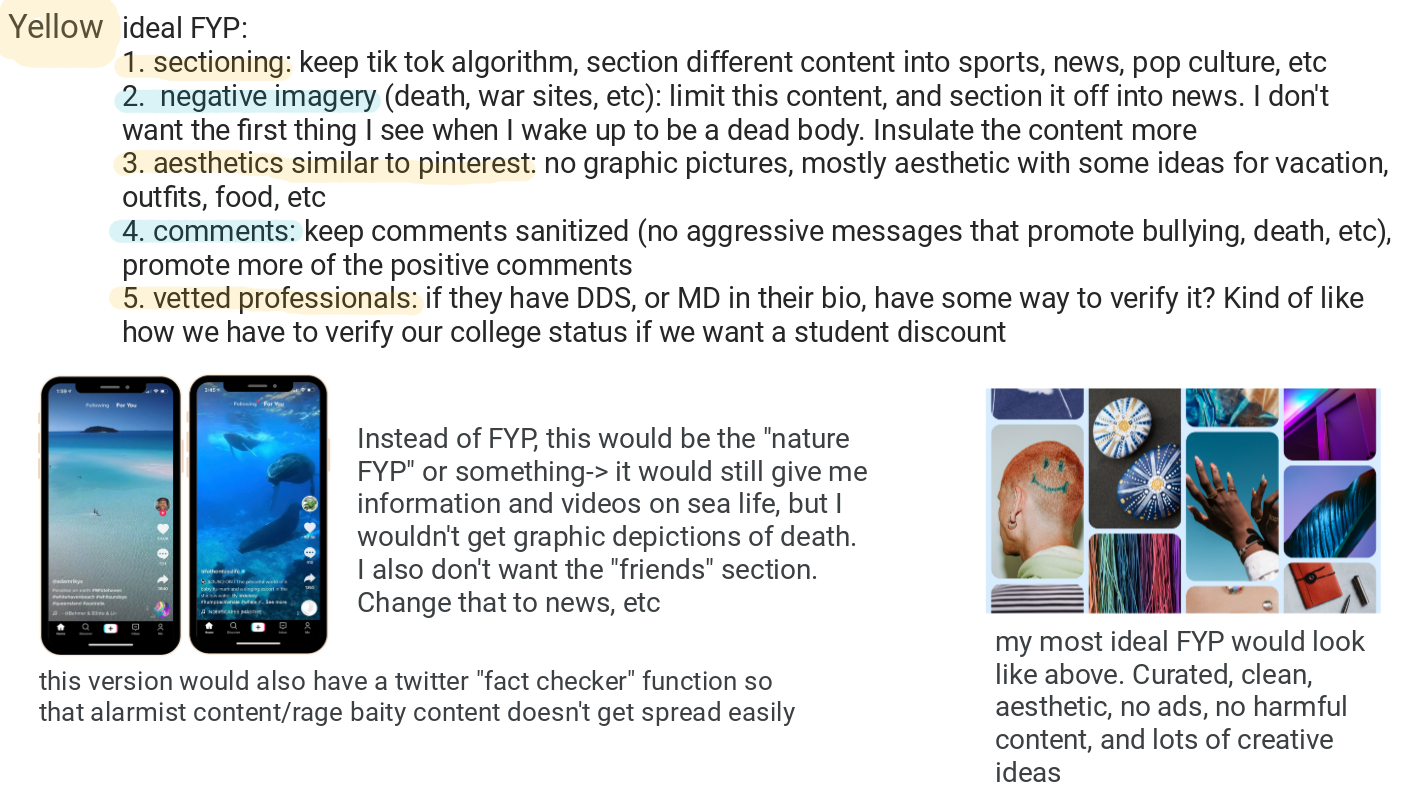}
            \caption{Design ideas from P16.}
            \label{fig:P16}
            \Description{An image of a digital interface design that outlines concepts for handling, sectioning, negative imagery, block layouts, comments, and creator verification.}
        \end{subfigure}
    \end{subfigure}
    \caption{Examples of participant prototypes.}
    \label{fig:designs}
    \Description{This figure contains several examples of the prototypes developed by participants. Each highlights details described further in the findings.}
\end{figure*}

\subsubsection{Contextualizing Engagement}
\label{subsub:context_engagement}
Participants' designs highlight a key tension: their contextual, intentional, and nuanced engagement was interpreted by platforms in simplistic, binary ways. Their designs aimed to create separation in three dimensions of intention: interest, amplification, and protection.

\textbf{Interest.} Many participants made design suggestions to refine the ``like'' button and comments to better capture users' intentions and interests. By separating engagement to reflect users' intentions around interest, participants aimed to combat the entanglement issues of ``Flattening’’ and ``Overloading’’ where they felt features controlled too many facets of the feed. For example, P03 suggested adding ``dislike'' options, but P11 noted that it could be ``not good for creators' mental health'' and that the algorithm tends to promote negative content. For comments, P01 proposed adding sentiment, which would prevent all comments from being interpreted as positive engagement.

\textbf{Amplification.} Participants wanted more nuanced ways to promote content to others. They distinguished between ``liking'' content and wanting to amplify, or see more of it in their feeds. Adding avenues for amplification divorced from interest aims to address ``Flattening’’ and ``Overloading’’ by disentangling user intentions and system actions. For instance, P01 added ``show more or less'' options to their design, explaining that it added dimensionality beyond ``like or dislike'' by indicating a desire to amplify content rather than indicating interest. Separating interest and amplification aims to address participants' complaints about mental health content, where it was of interest. Still, they do not always want it amplified, as it tends to overwhelm their feeds.

\textbf{Protection.} In adapting ``blocking'' features as seen in Figure \ref{fig:P12}, participants created ways to remove harmful content regardless of community guidelines to protect their mental well-being. They wanted the ability to block channels, keywords, or topics, with P08 wanting to ``throw out'' content and ``never have it come back''. These designs aim to mitigate the entanglement issues participants experienced with ``Guessing’’, ``Hiding’’, and ``Dangling’’, by creating clear removal avenues for harmful content. Rather than ``reporting'' which aligns with community guidelines, participants imagined softer protections for removing personally harmful content by ``flagging [content] that [is] personally upsetting or unwanted'' (P03) to ``protect against triggers'' (P10). Additional features included ``descriptions of why something might have been flagged'' (P10), ``a disclaimer if multiple people have flagged the content'' (P03), and ``auto-blur'' for triggering posts (P21).

\subsubsection{Consumption Control}
\label{subsub:consumption_control}
The second area of participant solutions revealed a need for interventions to support healthy content consumption and mental well-being; participants found current social media inventions missing or ineffective. Their designs created interventions that promoted intentional consumption through controlling engagement effects on feeds, intentional feed layouts, and personalized intervention strategies.

\textbf{Controlling Engagement Effects.} Participants wanted real-time control over their engagement to adjust feeds when they started having mood swings, which affected users' mental well-being. They envisioned sliders for real-time feed changes, allowing ``prioritized or deprioritized engagement'' (P05). Sliders to manage diversity, or ``how far'' (P05) recommendations deviate, were also suggested to control the appearance of content ``on the fringe'' (P09) of their interests. At the extremes, these sliders could allow engagement to be ``incognito'' (P04) and not affect their feeds. This direct control of engagement aims to address ``Hiding’’ and ``Disempowering’’ entanglement issues by exposing paths for user control.

\textbf{Intentional Layouts.} Participants criticized current social media layouts for being optimized for mindless consumption rather than mental well-being. Block layouts with ``aesthetic similar to Pinterest'' (P16) (see figure \ref{fig:P16}) and eliminating ``infinite scroll'' (P10) and auto-play, which participants noted encouraged ``doom scroll'' (P03), were proposed to create more intentional interactions by removing forced watch time. These designs aim to address the entanglement issues of ``Flattening’’, ``Disempowering’’, and ``Overloading’’ by introducing intentionality and reducing problematic engagement metrics.

\textbf{Personalized Interventions.} While participants found platform interventions aimed at content consumption, like ``stop scrolling'' message, potentially helpful, the lack of personalization made them ineffective. Participants wanted ``the ability to voluntarily turn on the warnings to take a break from social media'' (P09) with customizable conditions and messages, including condition logic for specific behaviors and message contents and presentation. Examples of some conditions included ``screen time'' that asked ``if the user wants to continue'' (P03) or limiting ``the number of posts that you see in a day'' (P06). For messaging, participants wanted ``more comprehensive mental health resources outside of the suicide hotline'' (P10). Personalized intervention designs aim to mitigate ``Hiding’’ and ``Disempowering’’ entanglements by making intervention features visible and controllable for users.

\subsubsection{Reclaiming Feed Control through Explicit Input}
\label{subsub:feed_control}
Frustration with unpredictable feed behaviors permeated participant designs, which emphasized user-centered, explicit controls. The designs introduced mechanisms to create contextual feeds by organizing and categorizing content to support their mental well-being.

\textbf{Organizing for Contextual Feeds.} Participants repeatedly requested separate feeds that allow for explicit organizational control over their feeds. These types of feeds were called ``containers'' (P04), ``groupings'' (P07), or ``multiple FY pages'' (P17), but shared similar functionality to create multiple feeds with fewer ``mood swings’’. For example, P17 wanted a mental health feed containing ``memes, personal experience, resources, etc'' in one dedicated place. Participants wanted to move content between feeds manually and ``toggle them on and off at will'' (P07) when the algorithm learned unwanted behaviors or included harmful content. Participants believed this design enabled them to create contextual interactions in their feeds and exercise greater control over content recommendations, thereby limiting issues such as disclosure leaks. Enabling explicit contextual control over interactions and feeds aims to address the entanglement issues of `Flattening’ and 'Disempowering’.

\textbf{Categorizing for Content Control.} Building on the previous design, participants wanted to provide ``more explicit input into what the algorithms show them'' (P12) through categorizing content with specific themes, or ``tags''. Tags, as seen in Figures \ref{fig:P09} and \ref{fig:P12}, would enable clear feedback and explicit control over harmful content, allowing users to add, remove, mute, or block content based on its theme. Several participants wanted filtering capabilities like Tumblr or AO3, with some requesting ``require tags on posts''. Most participants favored ``users curating their own tags'' (P07) as platform-generated tags could make ``communities might feel boxed-in'' (P07). Some participants thought about using tags as a starting point for content in the aforementioned feeds. Tags aim to address the entanglement issues of ``Guessing’’ and ``Hiding’’ by providing dedicated, transparent mechanisms for content control.

\section{Discussion}
\subsection{Theory Implication: Entanglement, Joined Systems, and Norman Design Interactions}
\label{dis:theory}
Our findings show that users experience entanglement, where they are unable to connect their actions to their consequences on social media platforms. Attempts to curate content for their mental well-being through interface interactions left participants entangled, exposing them to emotionally erratic feed behaviors that harmed their well-being. Participants heavily critiqued social media platforms and their emergent issues with entanglement, aspects of which have been explored in prior work on self-control and infinite scrolling feeds for platforms like YouTube~\cite{lukoff2021design}.

Our work reveals a fundamental design problem that we hypothesize has led to the current entanglement seen on social media platforms. Two distinct technological systems, each with its own function, have been merged into a single interface and platform (\eg, social networking and algorithmic curation systems in platforms like TikTok and Instagram). Classic examples of technological systems, such as Norman's doors and teapots~\cite{norman2013design}, provide clear, distinct functions that users can easily troubleshoot when problems arise; in contrast, modern algorithmic systems and their functions are often opaque. In many prior systems, such as adding search to email, integrating the system into the interface creates a new function clearly distinct from the original system. 

However, by integrating the distinct functions of social networking and algorithmic curation into a single, intertwined interface, current social media platforms obscure how the system works. Our results on ``Guessing'' and ``Dangling'' showed that traditional social media features (likes, comments, shares) now play entangled roles, serving both as platform interactions and as data for algorithmic personalization. Prior research shows that both social networking~\cite{feuston2019everyday,dyson2016systematic} and recommender systems~\cite{milton2022users} negatively affect users' mental well-being, further complicating the issue. The intertwined interfaces are evident in the features platforms offer, which have remained essentially unchanged after the integration of algorithmic curation, with only minimal additions, such as the ``not interested'' feature. Said plainly, the interface of social networking has been forced to control algorithmic curation on social media, leading to the entanglements we found in our study.

Our work on entanglement builds on the foundational work of~\citet{norman2013design} to investigate what happens when technological platforms, particularly those that affect users' behaviors and mental well-being, attempt to combine two systems. Successful navigation of Norman's action cycle requires a strong mental model of how a technology works. However, users cannot construct these mental models to reason about the actions to take to influence algorithmic behaviors. Overlaying recommender system affordances (\eg, implicit user preferences) onto traditional social media features (\eg, liking and commenting) obscures crucial information for building mental models, thereby creating entanglement.

The concept of entanglement can further help us understand and anticipate platform challenges arising from integrating different systems in the future. A timely example of where entanglement could be applied is generative AI in existing search engines. Search engines and generative AI differ in their processes and outcomes: one \textit{retrieves} information, while the other \textit{generates} it. Yet generative AI is currently integrated into existing search engines. We can already see that entanglement may be occurring in this case, with reports of issues with trust and cognitive damage~\cite {zhang2025generative,lee2025impact}. Entanglement could explain why distrust is occurring, as users struggle to distinguish between credible information from websites and that generated by generative AI. By using the proposed framework to analyze users' experiences interacting with the system, researchers can identify the specific issues users face, thereby disentangling the system to support users' needs and well-being.

\subsection{Design Implications: Combating Entanglement}
Entanglement disrupts users' ability to form mental models of the feeds by obscuring connections between actions and system response, directly affecting their mental well-being. Addressing entanglement requires rethinking the relationship between social media, recommender systems, users, and content. A straightforward solution to entanglement could be to map all desired features to buttons. However, that would over-encumber the interface, creating an equally stressful experience for users. We must find a balanced solution to support users without burdening them when interacting with these systems. 

Addressing entanglement requires a fundamental reconceptualization of how algorithmic curation supports users rather than ``one-size-fits-all’’ interface changes or algorithmic updates. While companies have financial incentives to keep users engaged on their platforms, users grow dissatisfied with platform design practices and are likely to migrate to alternative options. Thus, supporting users' needs ultimately should align with the platform's long-term interests. Some platforms have even begun implementing wellness features, such as watch notifications and activity centers, to promote well-being. In this section, we propose two design implications based on participants' designs (see Section~\ref{fig:designs}): restoring intentional control through contextually aware interfaces and reimagining algorithmic curation through a contextually aware lens.

\subsubsection{Restoring Intentional Control: ``Choose Your Own Adventure'' Interventions}
\label{dis:restore}
Our first design implication addresses entanglement issues by changing interfaces to establish context and communicate user intentions, thereby rebuilding the connections between users' actions and system responses. We frame these as ``choose your own adventure'' interventions, embodying users' control over their feeds to support their mental well-being.

\textbf{Personalized Safeguards for Mental Well-being.} 
Existing platform interventions for mental well-being, such as screen-time timers, ``stop scrolling'' messages, and crisis pop-ups, were ineffective for our participants. Participants perceived these interventions as generic and disingenuous, given their distrust of the platform's intentions. Moreover, in Section~\ref{sub:folk}, participants believed that these simplistic interventions removed their agency and choice in safeguarding their mental well-being.

We recommend a suite of user-tailored interventions that mitigate entanglement through restoring user agency and trust. For example, in Section~\ref{subsub:consumption_control}, participants envisioned customizable feed layouts and direct manipulation of feed behaviors to alter distressing patterns of behavior directly. Volume- or topic-based threshold or filter approaches were also suggested to allow mapping from users' plain language to interface behaviors, \eg, after 10 videos on a heavy topic, do not show any more for the rest of the day. Personalized interventions support users' agency in protecting their mental well-being.

These intervention approaches push against the ``one size fits all'' responses to extreme behaviors, instead prioritizing users' agency and early interventions to mediate harmful behavior patterns. However, implementing these interventions requires platforms to shift from viewing mental well-being as legal risk mitigation to genuine user support. Developing and evaluating diverse, well-grounded interventions requires joint research efforts between HCI and psychology to ensure they support mental health and well-being needs.

\textbf{Disentangling Context Through Separate Feeds.}
Current social media platforms are cross-contaminating user intentions and mental states by compressing them into single feeds. Participants designed modular feeds that respect users' diverse contexts, enabling them to maintain isolated feeds for distinct intentions, interests, and mental states.

We therefore recommend separate feeds that promote content organization. For example, platforms could provide multiple feeds, similar to browser ``tabs,'' allowing users to easily create, modify, and remove feeds on demand. These ``tabs'' could understand users' intentions through various methods, such as content ``tags'' (Section~\ref{subsub:feed_control}), aesthetic preferences, or explicit questions about emotional and mental states to inform feed creation. Platforms like BlueSky have adopted user-curated feeds that users can ``follow'' or manually curate. This approach effectively expands a user's ego-network beyond direct following to include users who contribute to those feeds. What our participants wanted differed from existing feeds: they wanted to retain the serendipitous discovery enabled by algorithmic curation while maintaining control over feed content.

This approach builds on participants' content organization designs from Section~\ref{subsub:feed_control}, which research shows supports autonomy and self-esteem~\cite{world2022world}. Implementing separate feeds requires maintaining separate user models and respecting contextual boundaries, which builds on prior calls for algorithmic feed control~\cite{simpson2022tame,milton2023see}. These designs must balance user control with the cognitive burden imposed by such systems. Evaluating these trade-offs through usability studies would need insights from psychology to assess mental health and well-being impacts.

\subsubsection{Reimagining Algorithmic Curation: From Accurate to Emotionally Dynamic}
Our second approach targets entanglement by shifting the curation algorithm's optimization from accuracy to contextual and emotional relevance. Current algorithms are seen as accurate but fail to account for context or user intentions.

\textbf{Content-Aware Recommendations.} Current algorithmic curation can not distinguish between different topics and tones of content that affect users' interactions and intentions in engagement. This lack of contextual awareness stems from joined systems (defined in Section~\ref{dis:theory}) as social media is inherently emotional~\cite{steinert2022emotions}, while recommender systems focus more on binary metrics. 

We propose an ensemble approach using multiple specialized models, each tailored to specific content types and contexts. Instead of a single monolithic model, multiple smaller models could be trained for specific contexts and user intentions. For example, separate models could be used for traditional recommendation contexts like ``BookTok'' content, and for more complex content like ``MentalHealthTok'' content that focuses on social support and mental well-being. An ensemble approach would support the multi-feed ``tabs'' suggested in Section~\ref{dis:restore}, with each tab powered by the appropriate models.  

The ensemble approach extends prior work on the contextual ambiguity in feeds~\cite{milton2023see,zhang2023pragmatic} and media richness theory~\cite{daft1986organizational}. Fully addressing contextual awareness requires collaboration between HCI, social computing, and recommender systems researchers to integrate users' insights about context into algorithmic design.

\textbf{User-Aware Recommendations.} 
Beyond content context, algorithmic curation must consider user context, particularly emotional and mental states, which determine users' capacity to engage with specific topics or tones. While recommender systems research has explored context~\cite{kulkarni2020context}, including location~\cite{fayyaz2020recommendation} and mood~\cite{polignano2021towards,deldjoo2020recommender}, mental state remains unexplored despite participants identifying it as distinct from mood or emotions. 

We propose integrating mental state through explicit user specification combined with system inference. For example, users could specify they prefer comforting content over emotionally heavy topics in the morning, allowing the system to adjust recommendations based on time, emotional capacity, and content weight. This approach explicitly incorporates the user's mental state into system behavior. However, modeling the complex interplay among mental state dimensions requires significant advances in user modeling technology and contextual inference. Conceptualizing and operationalizing mental state as a context is also a rich area for future research for psychology and HCI. Additionally, we need an understanding of how contextual aspects appear in interaction data and are digitized for algorithmic interpretation. User modeling and interaction analysis provide starting points for such work.

\section{Limitations and Future Work}
The limitations of our study primarily revolve around the participant pool. Our initial recruitment focused on university campuses despite attempts to reach a broader online audience. Thus, participants lean heavily toward college-aged adults, all of whom have at least some college education. Our participants were also skewed white and female, which continues the trend of struggling to recruit men and racial/ethnic minorities into studies for mental health~\cite{brown2014barriers,sanchez2020social}. Lastly, while we tracked whether participants reported experiencing mental health conditions (specifically depression, anxiety, and PTSD), we did not record which conditions or co-morbid diagnoses they specifically had out of respect for the participants' privacy. Participants were free to self-disclose during the workshops, but these conditions and individuals do not represent the entirety of the mental health community.

We still believe our findings open several possible research directions. Entanglement and the consequences on joined systems are a rich area of future research, allowing us to understand integrated platforms and identify issues that affect users' experiences and well-being. Considering the impacts of joined systems, this set a foundation for work evaluating current platforms through a new lens, updating existing knowledge and theories in social computing and HCI. Regarding design, our participants had concrete ideas for changing algorithmic curation to better support their mental well-being through features that enable explicit user control and mechanisms to organize recommended content. These ideas can be further studied in the areas of interface design and recommender systems when considering mental well-being. Expanding the research to more users, including those with other mental illnesses or marginalized identities, is vital to ensure designs are not over-generalized.

\section{Conclusion}
In this work, we examine how users experience algorithmic curation in social media and how they would like to see these systems evolve. Through design workshops, we discovered that users develop folk theories to make sense of their experiences with algorithmic curation. These folk theories showcase core issues with algorithmic design that we explain through entanglement, a phenomenon in which users cannot connect their actions to platform outcomes. We contribute to theory by building on Norman's action cycle to develop the entanglement framework, adding new dimensions to the gulfs of execution and evaluation, and reconceptualizing integrated systems. Our findings provide design implications that rethink current social media algorithmic curation to support users' mental well-being. 

\begin{acks}
We would like to thank Leah Ajmani and Mo Houtti for their invaluable feedback on this work. Funding for this research was partially provided by UL Research Institutes through the Center for Advancing the Safety of Machine Intelligence.
\end{acks}

\bibliographystyle{ACM-Reference-Format}
\bibliography{main}

\end{document}